\documentclass[10pt,a4paper,twocolumn]{article}
\usepackage{fukugo}
\usepackage{amsmath}
\usepackage{amsfonts}
\usepackage{amssymb}
\newcommand{\beq}{\begin{equation}}
\newcommand{\eeq}{\end{equation}}
\newcommand{\bea}{\begin{eqnarray}}
\newcommand{\eea}{\end{eqnarray}}

\begin{document}
%
\title{Distributed Spin in a String Picture of Hadrons}
%
%
%

\author{{\large{Akio SUGAMOTO}} \medskip \\
    \it{ Department of Physics and Chemistry,} \\
    \it{ Graduate School of Humanities and Sciences,       } \\
    \it{ Ochanomizu University                             }
         }
\date{  }
%
%
%
\maketitle
\pagestyle{empty}
\thispagestyle{empty}
\setlength{\baselineskip}{13pt}
%
%

%
Abstract
\hrulefill

In a string picture of hadrons, spins are distributed over the whole 
configuration of string.  According to this picture, spin of hadron 
is discussed in a dual gravity theory of QCD, towards realistic mass formulae 
of hadrons including hyperfine interactions.

\hrulefill

Recently, we studied the newly observed 
pentaquark baryons, consisting of four quarks and one anti-quark, in a string 
picture [BKST].  The non-perturbative behavior of 
quantum chromodynamics (QCD) was nicely taken into account by the 
dual gravity theory, or the AdS/CFT correspondence [M].  We 
used a QCD like model, having $N_{c}$ (=3) 
colored D4 branes and $N_{f}$ flavored D6 
branes [K]. To break the supersymmetry one spacial dimension along the D4 brane 
is compactified.  In this model simple mass formulae of 
pentaquarks were obtained, and the extreme smallness of their decay 
widths was qualitatively explained.  However, an important problem of how spin 
is incorporated in such a picture of hadrons, has not been solved there.

As is well known from the deep inelastic scattering experiments, only 
half of the momentum of nucleon is carried by quarks, but the remaining 
half is carried by gluons.  A similar but a more serious problem found from the 
same kind of experiment using  polarized electron and nucleon is that only 38 
\% of nuclear spin is carried by quarks. This is the famous "spin crisis" 
[Mak].  As a result, these experiments 
show that both momentum (or energy) and spin are distributed over the whole 
hadron.  

This is is a very good indication to support the string 
picture.  In this picture the identities of quarks or gluons 
disappear, and a hadron is a branched web of strings.  The web have also 
junctions where 
$N_{c}$ (=3) strings separate or merge.
The picture comes from the fact that we can assign two quantum numbers, color 
and flavor, on both ends of an open string which is the building block of 
hadrons, but the energy and the spin cannot be located locally.  It is well 
known that the energy is distributed on a string web, being proportional to its 
length.

As for the spin, the fermionic coordinate in superstring theory, 
$\psi^{\mu}(\tau, \sigma)$, a partner to the bosonic coordinate 
$X^{\mu}(\tau, \sigma)$, satisfies a commutation relation,
\begin{equation}
\{ \psi^{\mu}(\tau, \sigma), \psi^{\nu}(\tau, \sigma') 
\}=\eta^{\mu\nu}\delta(\sigma-\sigma'),
\end{equation}
so that $\frac{1}{\sqrt{2}}\psi^{\mu}(\tau, \sigma)$ becomes
the distribution function of $\gamma$ matrix, $\gamma^{\mu}$, over a 
string configuration.

The Gordon decomposition of the electromagnetic interaction reads
\begin{eqnarray}
S_{int}=-e/m\int d^4x 
\left(-\bar{\psi}(x)i\partial^{\mu}\psi(x)A_{\mu}(x) \right. 
\nonumber \\
\left.+\frac{1}{4}\bar{\psi}(x)\sigma^{\mu\nu}\psi(x)F_{\mu\nu}(x)\right).
\end{eqnarray}

This shows that a momentum density couples to the gauge field while a 
spin density couples to the field strength.  In terms of string this 
becomes
\begin{eqnarray}
& &-e\int d\tau d\sigma  F_{\mu\nu}(X) \nonumber \\
& &\times\left(\frac{\partial 
(X^{\mu},X^{\nu})}{\partial(\tau,\sigma)}+ 
4\pi\alpha' i \psi^{\mu}(\tau,\sigma)\tilde{\psi}^{\nu}(\tau,\sigma)\right).
\end{eqnarray}

Now we can understand that the "world sheet supersymmetry" 
\begin{eqnarray}
\left(\frac{\partial}{\partial\tau}-\frac{\partial}{\partial\sigma}\right)X^{\mu
}&\leftrightarrow & \sqrt{4\pi\alpha'}~\psi^{\mu}(\tau-\sigma) \\
\left(\frac{\partial}{\partial\tau}+\frac{\partial}{\partial\sigma}\right)X^{\mu
}&\leftrightarrow & \sqrt{4\pi\alpha'}~\tilde{\psi}^{\mu}(\tau+\sigma).
\end{eqnarray}
is a natural symmetry which interchanges the mometum (energy) dependent 
interaction and the spin dependent interaction in gauge theory.  Therefore, we 
adopt the "spinning string theory" having this  
world sheet supersymmetry for our purpose.

In a dual description of QCD, the background space is a curved space 
of (AdS-Schwarzshild$)_6 \times S^4$.  The main part of the metric is 
\begin{equation}
ds^2=f(u)(-dt^2+dz^2+d{\bf x}_{\perp}^2)+g(u)du^2+g(u)^{-1}d\vartheta^2,
\end{equation}
where $f(u)$ and $g(u)$ are given in [K], [BKST].

The spinning string in a curved space can be formulated, following 
the original paper by Iwasaki and Kikkawa [IK].
Accordingly, the action is given by
\begin{equation}
S=\int d\tau d\sigma \sqrt{-\mbox{det}~g_{ab}},
\end{equation}
with a world sheet metric
\begin{eqnarray}
g_{ab}&=&\partial_{a}X^{\mu}\partial_{b}X^{\nu}G_{\mu\nu}(x) \nonumber \\
&+&\frac{i}{2}\left(\bar{\psi}^{\mu}\hat{e}_{a}D_{b}\psi^{\nu}-D_{b}\bar{\psi}^{
\mu}\hat{e}_{a}\psi^{\nu}\right)G_{\mu\nu}(x),
\end{eqnarray}
where $\hat{e}_{a}=e_{a\bar{a}}\gamma^{\bar{a}}$ is the contraction 
of zwiebein by two dimensional $\gamma$ matrix, and
\begin{equation}
D_{a}\psi^{\mu}=\partial_{a}\psi^{\mu}+\partial_{a}X^{\nu}\Gamma^{\mu}_{\nu\lambda}(x)\psi^{\lambda}.
\end{equation}
Here we have to make some 
approximations. We are allowed to treate the spin perturbatively.  So, we 
expand the action in spin 
variables $\psi$, and we have $S=S_{(0)}+S_{(2)}+S_{(4)}+ \cdots$, where the 
number in the parenthesis shows the order in $\psi$.  The spin density is 
bilinear in $\psi$, so $S_{(4)}$ includes the hyperfine interaction or the 
spin-spin interaction.  The lowest action $S_{(0)}$ is the one studied in 
[BKST].  

Furthermore, the connection conditon of strings at 
junctions should be known.  This can be done using the world sheet 
supersymetry.  Namely, the balance of forces at a junction $\sum_{i=1-N_{c}} 
\partial_{\sigma}X^{\mu}_{(i)} \vert_{J}=0$ gives
\begin{equation}
\sum_{i=1-N_{c}} 
\left(\psi^{\mu}_{(i)}-\tilde{\psi}^{\mu}_{(i)}\right)\vert_{J}=0,
\end{equation}
while the ends points of $N_{c}$ strings termnating at a junction move with the 
same velocity, $\partial_{\tau}X^{\mu}_{(i)} \vert_{J}=v^{\mu}_{J} ~ 
(i=1-N_{c})$, leads to
\begin{equation}
\left(\psi^{\mu}_{(i)}+\tilde{\psi}^{\mu}_{(i)}\right)\vert_{J}=\lambda^{\mu}_{J
} ~~  
(i=1-N_{c}).
\end{equation}

Similar to the previous work [BKST], we first obtain 
a static solution of hadrons.  Then, solving the Schr\"odinger equation, or the 
Euler equation for a collective coordinate of translation, $R(t)$, the energy 
eigen-values of hadrons are obtained.  In this calculation we can fix the spin. 
 Since the total 
spin operator is known,
\begin{equation}
J^{\mu\nu}=\int d\sigma X^{[\mu}(\tau,\sigma)P^{\nu]}(\tau,\sigma)
+\frac{i}{2}\psi^{[\mu}(\tau,\sigma)\tilde{\psi}^{\nu]}(\tau,\sigma).
\end{equation}
In this way we can select the spin of   hadron, such as $J=\frac{1}{2}$ or 
$\frac{3}{2}$.  
We have to consider also the energy stored at junctions [I].  The details will be given elsewhere.

\vspace{5mm}

Acknowledgment
\hrulefill

The author acknowledges the helpful discussions and collaborations with Y. 
Imamura, M. Bando, T. Kugo and S. Terunuma.

\hrulefill

%

\end{document}